\documentstyle[12pt]{article}
\def\beq{\begin{equation}}
\def\eeq{\end{equation}}
\def\bea{\begin{eqnarray}}
\def\eea{\end{eqnarray}}
\def\bq{\begin{quote}}
\def\eq{\end{quote}}
\def\ve{\vert}
\def\nnb{\nonumber}
\def\ga{\left(}
\def\dr{\right)}
\def\aga{\left\{}
\def\adr{\right\}}

\def\rar{\rightarrow}
\def\nnb{\nonumber}
\def\la{\langle}
\def\ra{\rangle}
\def\nin{\noindent}
\begin{document}
\topmargin -1.5cm
\oddsidemargin +0.2cm
\evensidemargin -1.0cm
\pagestyle{empty}
\begin{flushright}
PM 95/05
\end{flushright}
\vspace*{5mm}
\begin{center}
\section*{ \bf QCD SPECTRAL SUM RULES \\ FOR HEAVY FLAVOURS$^{*)}$}
\vspace*{0.5cm}
{\bf S. Narison} \\
Laboratoire de Physique Math\'ematique\\
Universit\'e de Montpellier II\\
Place Eug\`ene Bataillon\\
34095 - Montpellier Cedex 05\\
\vspace*{1.5cm}
{\bf Abstract} \\ \end{center}
\vspace*{2mm}
\noindent
Recent developments in the uses of QCD spectral sum rules (QSSR)
for heavy flavours are summarized and updated.
QSSR results are compared with the existing data and with the ones from
alternative approaches.
\vspace*{4.0cm}
\noindent
\rule[.1in]{16.0cm}{.002in}

\noindent
$^{*)}$ Talk given at the Cracow Epiphany Conference
on Heavy Quarks, 6th January (1995), in honour of the 60th birthday of
Kacper Zalewski (to appear in Acta Physica Polonica). This is
an updated version of the talks given  by the author at
the {\it XXIXth  Rencontre de Moriond}, M\'{e}ribel (1994),
CERN-TH.7277/94 (1994) and at the {\it QCD94 Conference}, Montpellier
(1994)
CERN-TH.7444/94 (1994). An extended version of this review will be
published
in {\it Recent Developements of QCD spectral sum rules} by World
Scientific
Company.
\vspace*{0.5cm}
\noindent


\begin{flushleft}
PM 95/05\\
March 1995
\end{flushleft}
\vfill\eject
\pagestyle{empty}
\setcounter{page}{1}
\pagestyle{plain}

\def\beq{\begin{equation}}
\def\eeq{\end{equation}}
\def\bea{\begin{eqnarray}}
\def\eea{\end{eqnarray}}
\def\bq{\begin{quote}}
\def\eq{\end{quote}}
\def\ve{\vert}
\parskip 0.3cm

\def\nnb{\nonumber}
\def\ga{\left(}
\def\dr{\right)}
\def\aga{\left\{}
\def\adr{\right\}}

\def\rar{\rightarrow}
\def\nnb{\nonumber}
\def\la{\langle}
\def\ra{\rangle}
\def\nin{\noindent}
\def\alf{\alpha_s}
\section*{\bf QCD SPECTRAL SUM RULES\\
 FOR HEAVY FLAVOURS\footnote{
This is
an updated version of the talks given  by the author at
the {\it XXIXth  Rencontre de Moriond}, M\'{e}ribel (1994),
CERN-TH.7277/94 (1994) and at the {\it QCD94 Conference}, Montpellier
(1994),
CERN-TH.7444/94 (1994). An extended version of this review will be
published
in {\it Recent Developements of QCD spectral sum rules} by World
Scientific
Company.}}
\vspace*{0.5cm}
{\bf S. Narison} \\
\vspace{0.2cm}
\nin
Laboratoire de Physique Math\'ematique,
Universit\'e de Montpellier II\\
Place Eug\`ene Bataillon,
34095 - Montpellier Cedex 05\\
\vspace{1cm}
\nin
Recent developments in the uses of QCD spectral sum rules (QSSR)
for heavy flavours are summarized and updated.
QSSR results are compared with the existing data and with the ones from
alternative approaches.
\section{Introduction} \par
\nin
We have been living with QCD spectral sum rules (QSSR)
(or QCD sum rules, or ITEP sum rules, or hadronic sum rules...)
for 15 years, within the impressive ability
of the method for describing the complex phenomena of hadronic
physics with the few universal ``fundamental" parameters of the QCD
Lagrangian
(QCD coupling $\alpha_s$, quark masses
and  vacuum condensates built from the quarks
and/or gluon fields, which parametrize the
non-perturbative phenomena). The approach might be very close to the
lattice calculations as it also uses the first principles of QCD, but
unlike the case of the lattice, which is based  on
sophisticated numerical simulations, QSSR is
quite simple as it is a semi-analytic approach based
on a semiperturbative expansion and Feynman graph
techniques implemented in an Operator Product Expansion (OPE),
where  the condensates contribute as higher-dimension
operators. The QCD information is transmitted to the data
via a dispersion relation obeyed by the hadronic correlators,
in such a way that
in this approach, one can {\it really control and in some sense localize}
the origin of the numbers obtained from the analysis. With this
simplicity, QSSR can describe in an elegant way the complexity
of the hadron phenomena, without waiting for a complete understanding
of the confinement problem.

\nin
One can fairly say that QCD spectral sum rules already started, before
QCD, at the time of current algebra, in 1960, when different {\it ad hoc}
superconvergence sum rules, especially the Weinberg and
Das--Mathur--Okubo sum rules, were proposed but they came
under control only with the advent of QCD \cite{FLO}. However,
the main flow comes from the classic paper of
Shifman--Vainshtein--Zakharov
\cite{SVZ} (hereafter referred to as SVZ), which goes beyond
the {\it na\"\i ve} perturbation
theory thanks to the inclusion of the vacuum condensate effects in
the OPE (more
details and more complete discussions of QSSR and its various
applications to hadron physics can be found, for instance, in
\cite{SNB}).

\nin
In this talk,
I shall present aspects of QSSR in the analysis of the
properties of heavy flavours. As I am limited in space-time,
I cannot cover in detail here all QSSR applications to the
heavy-quark physics. I will only focus on the following
topics, which I think are important in the development of the
understanding of the heavy-quark properties in connection with the
progress done recently in the heavy quark effective theory (HQET)
(or infinite mass effective theory (IMET)) and in lattice calculations:

\nin
-- the heavy-quark-mass values,
\nin
--the meson-quark mass difference and the heavy quark kinetic energy,

\nin
-- the pseudoscalar decay constants,
\nin
-- the bag constants and the CP-violation parameters,

\nin
-- the heavy to light exclusive decays,

\nin
-- slope of the Isgur-Wise (IW) function and determination of $V_{cb}$,

\nin
-- properties of the hybrids and $B_c$-like hadrons.

 \section{ QCD spectral sum rules}
\nin
In order to illustrate the QSSR method in a pedagogical way, let us
consider the two-point correlator:
\bea
\Pi^{\mu\nu}_b &\equiv& i \int d^4x ~e^{iqx} \
\la 0\vert {\cal T}
J^{\mu}_b(x)
\ga J^{\nu}_b(o)\dr ^\dagger \vert 0 \ra \nnb \\
&=& -\ga g^{\mu\nu}q^2-q^\mu q^\nu \dr \Pi_b(q^2,M^2_b),
\eea
where $J^{\mu}_b(x) \equiv \bar b \gamma^\mu b (x)$ is the local vector
current of the $b$-quark. The correlator obeys the well-known
K\"allen--Lehmann dispersion relation:
\beq
\Pi_b(q^2,M^2_b) = \int_{4M^2_b}^{\infty} \frac{dt}{t-q^2-i\epsilon}
{}~\frac{1}{\pi}~\mbox{Im}  \Pi_b(t) ~ + ...,
\eeq
where ... represent subtraction points. This $sum~rule$
expresses in a clear way the {\it duality} between the spectral
function Im$ \ \Pi_b(t)$, which can be measured experimentally, as here
it is related to the $e^+e^-$ into $\Upsilon$-like states total
cross-section, while $\Pi_b(q^2,M^2_b)$ can be calculated directly in
QCD, even at $q^2=0$,
thanks to the fact that $M^2_b-q^2 \gg \Lambda^2$.
The QSSR is an improvement on the previous
dispersion relation.

\nin
On the QCD side, such an improvement is achieved by adding
to the usual perturbative expression of the correlator,
the non-perturbative contributions as parametrized by the vacuum
condensates of higher and higher dimensions in the OPE \cite{SVZ}:
\bea
\Pi_b (q^2,M^2_b)
&\simeq& \sum_{D=0,2,4,...}\frac{1}{\ga M^2_b-q^2 \dr^{D/2}} \nnb
\\
&.&\sum_{dim O=D} C^{(J)}(q^2,M^2_b,\nu)\la O(\nu)\ra, \nnb \\
&&
\eea
where $\nu$ is an arbitrary scale that separates the long- and
short-distance dynamics; $C^{(J)}$ are the Wilson coefficients calculable
in perturbative QCD by means of Feynman diagrams techniques:
$D=0$
corresponds to the case of the na\"\i ve perturbative contribution;
$\la O \ra$ are
the non-perturbative condensates built from the quarks or/and gluon
fields. For $D=4$, the condensates that can be formed are the
quark $M_i \la\bar \psi \psi \ra$ and gluon $\la\alpha_s G^2 \ra$
ones; for
$D=5$, one can have the mixed quark-gluon condensate $\la\bar \psi
\sigma_{\mu\nu}\lambda^a/2 G^{\mu\nu}_a \psi \ra$, while for $D=6$
one has, for instance, the
triple gluon $gf_{abc}\la G^aG^bG^c \ra$ and the four-quark
$\alpha_s \la \bar \psi \Gamma_1 \psi \bar \psi \Gamma_2 \psi \ra$,
where $\Gamma_i$ are generic notations for any Dirac and colour matrices.
The validity of this expansion has been understood formally, using
renormalon techniques (IR renormalon ambiguity is absorbed into the
definitions of the condensates)
\cite{MUELLER} and by building  renormalization-invariant
combinations of the condensates (Appendix of \cite{PICH} and references
therein). The SVZ expansion is phenomenologically confirmed from the
unexpected
accurate determination of the QCD coupling $\alpha_s$
and from the measurement of the QCD condensates from semi-inclusive
tau decays and spectral moments \cite{PICH,ALFA}. In the present case
of heavy-heavy
correlators the OPE is much simpler, as one can show \cite{HEAVY}-
\cite{BC}
that the heavy-quark condensate
effects can be included into those of the gluon condensates, so
that, up to $D\leq 6$, only the $\la \alpha_sG^2\ra$ and $g\la G^3\ra$
condensates appear in
the OPE. Indeed, SVZ have, originally, exploited this feature for their
first estimate of the gluon condensate value, though the validity of
their result has been criticized later on
\cite{HEAVY,BELL}-\cite{DODO}.

\nin
For the phenomenological side, the improvement comes from the uses of
either a finite number of derivatives and finite values of $q^2$
(moment sum rules):
\bea
{\cal M}^{(n)}& \equiv& \frac{1}{n!}\frac{\partial^n \Pi_b(q^2)}
{\ga \partial q^2\dr^n} \Bigg{\vert} _{q^2=0} \nnb \\
&=& \int_{4M^2_b}^{\infty} \frac{dt}{t^{n+1}}
{}~\frac{1}{\pi}~ \mbox{Im}  \Pi_b (t),
\eea
or an infinite number of derivatives and infinite values of $q^2$, but
keeping their ratio fixed as $\tau \equiv n/q^2$
(Laplace or Borel or exponential sum rules):
\beq
{\cal L}(\tau,M^2_b)
= \int_{4M^2_b}^{\infty} {dt}~\mbox{exp}(-t\tau)
{}~\frac{1}{\pi}~\mbox{Im} \Pi_b(t).
\eeq
There also exist non-relativistic versions of these two sum rules,
which are convenient quantities to work with in the large-quark-mass
limit. In these cases, one introduces non-relativistic
variables $E$ and $\tau_N$:
\beq
t \equiv (E+M_b)^2 \ \ \ \ \mbox{and} \ \ \ \  \tau_N = 4M_b\tau .
\eeq
In the previous sum rules,
the gain comes from the weight factors, which enhance the
contribution of the lowest ground-state meson to the spectral integral.
Therefore, the simple duality ansatz parametrization:
\beq
``\mbox{one resonance}"\delta(t-M^2_R)
 \ + \
 ``\mbox{QCD continuum}" \Theta (t-t_c),
\eeq
of the spectral function,
gives a very good description of the spectral integral, where the
resonance enters via its coupling to the quark current. In the case
of the $\Upsilon$, this coupling can be defined as:
\beq
\la 0\vert \bar b\gamma^\mu b \vert \Upsilon \ra =
 \sqrt{2} \frac{M^2_\Upsilon}
{2\gamma_\Upsilon}.
\eeq
The previous
feature
has been tested in the light-quark channel from the $e^+e^- \rar$
$I=1$ hadron data and in the heavy-quark ones from the
$e^+e^- \rar \Upsilon$ or $\psi$ data, within a good
accuracy.
To the previous sum rules, one can also add the ratios:
\beq
{\cal R}^{(n)} \equiv \frac{{\cal M}^{(n)}}{{\cal M}^{(n+1)}}~~~~~
\mbox{and}~~~~~
{\cal R}_\tau \equiv -\frac{d}{d\tau} \log {{\cal L}},
\eeq
and their finite energy sum rule (FESR) variants, in order to fix
the squared mass of the ground state.
\nin
In principle, the pairs $(n,t_c)$, $(\tau,t_c)$ are free external
parameters in the analysis, so that the optimal result should be
insensitive to their variations. Stability criteria, which are equivalent
to the variational method, state that the best results should
be obtained at the minimas or at the inflexion points in $n$ or $\tau$,
while stability in $t_c$ is useful to control the sensitivity of the
result in the changes of $t_c$ values. To these stability criteria are
added constraints from local duality FESRs, which
correlate the $t_c$ value to those of the ground state mass and
coupling \cite{FESR}. Stability criteria have also been tested in
models such as the
harmonic oscillator, where the exact and approximate
solutions are known \cite{BELL}. The {\it most conservative
optimization criteria}, which include various types of optimizations
in the literature, are the following:
\nin
the $optimal$ result is obtained in the region, starting at
the beginning of $\tau / n$ stability (this corresponds in most
of the cases to the so-called plateau often discussed in the literature,
but in my opinion, the interpretation of this nice plateau as a
sign of a good continuum model is not sufficient, in the sense
that the flatness of the
 curve extends in the uninteresting high-energy region where the
properties of the ground state are lost),
until the beginning of the $t_c$
stability, where the value of $t_c$ more or less
corresponds in some cases to the one fixed by FESR duality constraints.
\nin
The earlier {\it sum rule window} introduced by SVZ, stating that the
optimal result should be in the region where both the non-perturbative
and continuum contributions are {\it small}, is included in the previous
region.
 Indeed, at the stability
point, we have an equilibrium between the continuum and non-perturbative
contributions, which are both small,
while the OPE is still convergent  at this point.
\section{The heavy-quark-mass values}
\subsection{The running masses}
\nin
Here, we will summarize the recent results obtained in \cite{SNM}, where
an improvement and an update of the existing results have been done,
with the emphasis that the apparent discrepancy encountered in the
literature is mainly due to the different values of $\alpha_s$ used by
various authors. Using the $world~average$
value $\alpha_s(M_Z)=0.118 \pm 0.006$ \cite{BETHKE,PDG,HIN}and a
$conservative$
value$\la \alpha_s G^2\ra= (0.06\pm 0.03)$ GeV$^4$ \cite{SNB,PICH},
the {\it first direct
determination} of the running mass to two loops  in the
$\overline{MS}$-scheme, from the $\Psi$ and
$\Upsilon$ systems, is \cite{SNM}:
\bea
\overline{m}_c(M^{PT2}_c) &=& (1.23 ^{+ 0.02}_{-0.04}\pm 0.03
)~ \mbox{GeV} \nnb \\
\overline{m}_b(M^{PT2}_b)
&=& (4.23 ^{+ 0.03}_{-0.04}\pm 0.02)~ \mbox{GeV},
\eea
where the errors are respectively due to $\alpha_s$ and to the gluon
condensate.
\nin
Using the previous result in (10) and the expression of the running
mass to two-loops \cite{FLO,SNB}:
\beq
\overline{m}_Q(\nu) =  \hat{m}_Q\ga -\beta_1\frac{\alf(\nu)}{\pi}\dr
^{-\gamma_1/\beta_1}
\times\aga 1~+~\frac{\beta_2}{\beta_1}
\ga \frac{\gamma_1}{\beta_1}-\frac{\gamma_2}{\beta_2}\dr
\ga \frac{\alf}{\pi}\dr
\adr ,
\eeq
in terms of the invariant mass
$\hat{m}_Q$, one can extract the running mass at another scale;
$\gamma_1=2$ and $\gamma_2= 101/12-5n_f/18$, $\beta_1=-11/2+n_f/3,~
\beta_2=-51/4=19n_f/12$ are the mass anomalous dimensions and the
$\beta$-function in the $\overline{MS}$-scheme. Then,
 one obtains at 1 GeV:
\bea
\overline{m}_c(1~\mbox{GeV}) &=& (1.46 ^{+ 0.09}_{-0.05}\pm 0.03
)~ \mbox{GeV} \nnb \\
\overline{m}_b(1~\mbox{GeV})
&=& (6.37 ^{+ 0.64}_{-0.39}\pm 0.07)~ \mbox{GeV},
\eea
By combining the previous value of the running $b$-quark mass
 with the $s$-quark one evaluated at 1 GeV, which we take in the range:
 $\overline{m}_s$(1 GeV)= 150-230 MeV
 \cite{SN1,GASSER},
 one obtains the scale-independent ratio:
\beq
\overline {m}_b/\overline {m}_s \simeq 33.5 \pm 7.6,
\eeq
a result of great interest for model-building and SUSYGUT-phenomenology.
\subsection{The pole masses}
\nin
One can transform the results on the running masses
into the $perturbative$ pole masses by using the perturbative
relation \cite{SN1}:
\beq
M_Q(\nu) = \overline{m}_Q(\nu)\Big\{1+
\ga\frac{\alpha_s}{\pi}\dr\ga\frac{4}{3}+
2\ln{\frac{\nu}{M_Q}}\dr +...\Big\},
\eeq
where the constant term of
the $(\alpha_s/\pi)^2$ is known to be: $K_b\simeq 12.4,~K_c\simeq 13.3$
\cite{GRAY}.Then, we obtain, to two-loop accuracy:
\bea
M^{PT2}_c & = & (1.42 \pm 0.03
)~ \mbox{GeV} \nnb \\
M^{PT2}_b
&= &  (4.62 \pm 0.02)~ \mbox{GeV}.
\eea
It is informative to compare these values with the ones from the pole
masses from non-relativistic sum rules to two loops:
\bea
M^{NR}_c & =& (1.45^{+ 0.04}_{-0.03} \pm 0.03
)~ \mbox{GeV} \nnb \\
M^{NR}_b
& =&  (4.69 _{-0.01}^{+0.02}\pm 0.02)~ \mbox{GeV}.
\eea
 A similar comparison
can be done at three-loop accuracy. One obtains:
\bea
M^{PT3}_c  & =&  (1.62 \pm 0.07 \pm 0.03
)~ \mbox{GeV} \nnb \\
M^{PT3}_b
&=&  (4.87 \pm 0.05 \pm 0.02)~ \mbox{GeV},
\eea
to be compared with
 the $dressed~mass$:
\beq
  M_b^{nr}= (4.94\pm 0.10 \pm 0.03)~\mbox{ GeV},
\eeq
obtained from
a non-relativistic Balmer formula based on a $\bar bb $ Coulomb
potential and including higher order $\alpha_s
$-corrections \cite{YND}.
\nin
One can remark that the radiative $\alpha_s^2$ correction
is large and causes a positive shift of about 250 MeV on the value of
the pole mass $M_b$.
One can also remark that at the two and three loop-accuracies,
the mass-difference between the relativistic
and non-relativistic pole masses is about 70 MeV.
The interpretation of this mass-difference is not quite
well understood. If one
has in mind that the non-relativistic pole mass contains a
non-perturbative piece due to Coulombic interactions, which can be of
the same origin as
the one induced by the truncation of the perturbative series at large
order, then one can consider this value as a phenomenological
estimate of the renormalon contribution, which is comparable
in strength with the estimate of about 100-133 MeV from the summation of
higher order corrections of large order perturbation theory
\cite{BENEKE}.
\nin
An extension of the previous analysis of the
$\Psi$ and $\Upsilon$-systems to the case of the $B$ and $B^*$
mesons leads to the value:
\beq
 M_b^{PT2}=(4.63 \pm 0.08) ~\mbox{ GeV},
\eeq
in good agreement with the previous results, but less accurate.
\subsection{The $b$ and $c$ pole-mass-difference}
\nin
One can also use the previous results, in order to deduce the
mass-difference between the $b$ and $c$ (non)-relativistic pole masses:
\beq
M_b(M_b)-M_c(M_c) = (3.22 \pm 0.03)~ \mbox{GeV},
\eeq
in good agreement (within the errors) with potential model expectations
\cite{RICH,PDG}. A direct comparison of this mass-difference with the one
from the analysis of the inclusive $B$-decays needs however a better
understanding of the mass definition and of the value of the scale
entering into these decay-processes. Indeed, if one chooses to evaluate
these pole masses at the scale $\nu=M_b$, which can be a natural scale
for this process, one obtains to two-loop accuracy:
\beq
M_c(\nu=M_b)= (1.08 \pm 0.04)~\mbox{GeV},
\eeq
which leads to the mass-difference:
\beq
M_b-M_c|_{\nu=M_b}= (3.54 \pm 0.05)~\mbox{GeV},
\eeq
in good agreement with the one extracted from the analysis
of the inclusive $B$-decays \cite{VAIN}.
\section{The meson-quark-mass gap and the heavy-quark-kinetic energy}
\nin
The meson-quark mass gap  $\bar \Lambda$ is in important input
in HQET (IMET) approach. It can be defined
as \cite{HQET}
\footnote{We are aware of the
fact that in the lattice calculations, $\bar \Lambda$ defined in this
way can be affected by
renormalons \cite{LATT}.}:
\beq
M_B=M_b+\bar \Lambda -\frac{1}{2M_b}\ga K+3\Sigma \dr,
\eeq
where:
\beq
K= \frac{1}{2M_B} \la B(v)|{\cal K}|B(v) \ra
{}~~~~~~\mbox{and}~~~~~~
\Sigma= \frac{1}{6M_B} \la B(v)|{\cal S}|B(v) \ra
\eeq
correspond respectively to the matrix elements of the kinetic and of
the chromomagnetic operators:
\beq
{\cal K}\equiv \bar h (iD)^2 h~~~~~~~
\mbox{and}~~~~~~{\cal S}\equiv\frac{1}{2}\bar h \sigma_{\mu \nu}
 F^{\mu \nu} h,
\eeq
where $h$ is the heavy quark field and $F^{\mu \nu}$ the electric field
tensor.
\nin
The estimate of $\bar \Lambda$ from HQET-sum rules leads to \cite{SNH}:
\beq
\bar \Lambda  \simeq (0.52 -0.70)~\mbox{GeV},
\eeq
in good agreement with the previous results \cite{NEUB,NEU}, though less
accurate as we have taken a larger range of variation for the continuum
energy.
An anologous sum rule in the full QCD theory leads to \cite{SN4}:
\beq
\bar \Lambda  \simeq (0.6 -0.80)~\mbox{GeV},
\eeq
which combined together leads to the intersecting range of values
\cite{SNH}:
\beq
\bar \Lambda  \simeq (0.65 \pm 0.05)~\mbox{GeV}.
\eeq
The sum rule estimate of the kinetic energy gives \cite{SNH}:
\beq
K\simeq -(0.5 \pm 0.2) ~\mbox{GeV}^2
\eeq
where the large error, compared with the previous result of \cite{BRAUN},
is due to the absence of the stability point with
respect to the variation of the continuum energy threshold. By combining
the previous estimates with the one of the chromomagnetic energy:
\beq
\Sigma \simeq \frac{1}{4}(M^2_{B^*}-M^2_B),
\eeq
one deduces the value of the pole mass to two-loop accuracy:
\beq
M_b=(4.61 \pm 0.05)~\mbox{GeV},
\eeq
in good agreement with the previous values from the sum rules in the full
theory and (within the errors) with the earlier HQET results of
\cite{NEUB}
\section{The pseudoscalar decay constants}
\subsection{Estimate of the decay constants}
\nin
The decay constants $f_P$ of a pseudoscalar meson $P$ are defined as:
\beq
(m_q+M_Q)\la 0\vert \bar q (i\gamma_5)Q \vert P\ra
 \equiv \sqrt{2} M^2_P f_P,
\eeq
where in this normalization $f_\pi = 93.3$ MeV.
A {\it rigorous}
upper bound on these couplings can be derived from the
second-lowest superconvergent moment:
\beq
{\cal M}^{(2)} \equiv \frac{1}{2!}\frac{\partial^2 \Psi_5(q^2)}
{\ga \partial q^2\dr^2} \Bigg{\vert} _{q^2=0},
\eeq
where $\Psi_5$ is the two-point correlator associated to the pseudoscalar
current. Using the positivity of the higher-state contributions to the
spectral function, one can deduce \cite{SNZ}:
\beq
f_P \leq \frac{M_P}{4\pi} \aga 1+ 3 \frac{m_q}{M_Q}+
0.751 \bar{\alpha}_s+... \adr,
\eeq
where one should not misinterpret the mass-dependence in this
expression compared to the one expected from heavy-quark symmetry.
Applying this result to the $D$ meson, one obtains:
\beq
f_D \leq 2.14 f_\pi .
\eeq
Although
presumably quite weak, this bound, when combined with the recent
determination to two loops \cite{SN2}:
\beq
\frac{f_{D_s}}{f_D} \simeq (1.15 \pm 0.04)f_\pi ,
\eeq
implies
\beq
f_{D_s} \leq (2.46 \pm 0.09)f_\pi ,
\eeq
which is useful for a comparison with the recent measurement of $f_{D_s}$
from WA75: $f_{D_s} \simeq (1.76 \pm 0.52)f_\pi$ and from CLEO:
$f_{D_s} \simeq (2.61 \pm 0.49)f_\pi$.
One cannot push, however, the uses
of the moments to higher $n$ values in this $D$ channel, in order to
minimize the continuum contribution to the sum rule with the aim to
derive an estimate of the decay constant because the QCD series
will not converge at higher $n$ values.
In the $D$ channel, the most appropriate sum rule is the
Laplace sum rule. The results from different groups are consistent
for a given value of the $c$-quark mass. Using the table in
\cite{SN2} and the value of the perturbative pole mass obtained
previously, one obtains to two loops:
\beq
f_D \simeq (1.35 \pm 0.04\pm 0.06)f_\pi~~~~~~~ \Rightarrow ~~~~~~~
f_{D_s} \simeq (1.55 \pm 0.10)f_\pi .
\eeq
For the $B$ meson, one can either work with the Laplace, the
moments or
their non-relativistic variants. Given the previous value of $M_b$, these
different methods give consistent values of $f_B$, though the one
from the non-relativistic sum rule is very inaccurate due to the huge
effect of the radiative corrections in this method. The best value comes
from the Laplace sum rule; from the table in \cite{SN2}, one
obtains:
\beq
f_B \simeq (1.49\pm 0.06\pm 0.05)f_\pi ,
\eeq
while \cite{SN2}:
\beq
 \frac{f_{B_s }}{f_B}\simeq 1.16 \pm 0.04,
\eeq
where the most accurate estimate comes from the ``relativistic" Laplace
sum rule.
\nin
The apparent disagreement among different existing
QSSR numerical results in the literature is
$not$ essentially due to the choice of the continuum threshold as
misleadingly
claimed in the literature $but$ is mainly due to the
different values of the quark masses used because the decay constants
are very sensitive to that quantity as shown explicitly in \cite{SN2}.
\subsection{Static limit and 1/M-corrections to $f_B$}
\nin
One could notice, since the {\it first} result
$f_B \simeq f_D$ of \cite{SN3}, a large violation of the
scaling law expected from heavy-quark symmetry.
 This is due to the large 1/$M_b$-correction
 found from the HQET sum rule \cite{NEU} and from the one in full
QCD \cite{SN4,SNA}. Using the estimate of the decay constant in the
static limit \cite{SNH}:
\beq
f_B^{\infty} \simeq (1.98 \pm 0.31)f_\pi,
\eeq
and the previous estimates of $f_B$ and $f_D$ in the full theory, the
quark-mass dependence of the decay constant can be parametrized as:
\beq
f_B \sqrt{M_b} \simeq (0.33 \pm 0.06)~\mbox{GeV}^{3/2}
\alpha_s^{1/{\beta_1}}
.\aga 1-\frac{2}{3}\frac{\alpha_s}{\pi}-
\frac{A}{M_b}+\frac{B}{M^2_b}\adr,
\eeq
by including the quadratic mass corrections, where:
\beq
A \approx 1.1~ \mbox{GeV}~~~~~\mbox{and}~~~~~
B \approx 0.7~\mbox{GeV}^2,
\eeq
while a linear parametrization
leads to:
\beq
A \simeq (0.6\pm 0.1) ~\mbox{GeV},
\eeq
in accordance with previous findings \cite{NEU,SN4,SNA} and with the
lattice results \cite{PENE}.
\nin
One can $qualitatively$ compare this result with the one obtained from
the
analytic expression of the moment or from the
semilocal duality sum rule, which leads to the $interpolating~formula$
\cite{ZAL}:
\beq
f_B \sqrt{M_b} \approx  \frac{E^{3/2}_c}{\pi} \alpha_s^{1/{\beta_1}}
\ga \frac{M_b}{M_B} \dr^{3/2}\Bigg\{ 1-\frac{2}{3}\frac{\alpha_s}{\pi}
+\frac{3}{88}\frac{E^2_c}{M^2_b}
-\frac{\pi^2}{2} \frac{\la \bar uu \ra}{E^3_c}+...\Bigg\} ,
\eeq
and gives:
\bea
A &\approx &\frac{3}{2}(M_B-M_b) \simeq 1~\mbox{GeV},\nnb \\
B &\approx& \frac{3}{88}E_c^2-\frac{9}{8}(M_B-M_b)^2 \simeq 0.5~\mbox
{GeV}^2,
\eea
in agreement with the previous numerical estimate.
\section{The bag constants and the CP-violation parameters}
\subsection{Estimate of the bag constant $B_B$}
\nin
The $B^0$-$\bar{B}^0$ mixing is gouverned by the $B_B$-parameter as:
\beq
\la \bar{B}^0 | \bar b\gamma_\mu^L d \bar b\gamma_\mu^L d |B^0 \ra =
\frac{4}{3} f^2_BM^2_BB_B(\nu),
\eeq
where one can introduce the invariant bag parameter $\hat{B}_B$ as:
\beq
\hat{B}_B \equiv B_B(\nu) (\alpha_s(\nu))^{-6/23}.
\eeq
We have tested the validity
of the vacuum saturation $B_B=1$ of the bag constant, using a
sum rule analysis of the four-quark two-point correlator to two loops
\cite{PIVO} following the leading order work of \cite{BB}. We
found that the radiative corrections due to the non-factorizable
contributions are quite small. Under some
physically reasonable assumptions for the spectral function, we found
that the vacuum saturation estimate is only violated by about $15\%$,
giving:
\beq
 B_B \simeq 1 \pm 0.15.
\eeq
By combining this result with the one for $f_B$, we deduce:
\beq
f_B\sqrt{B_B} \simeq  (197\pm 18)~{\mbox MeV},
\eeq
if we use the normalization $f_\pi= 132$ MeV, which is $\sqrt 2$
times the one defined in (30),
in excellent agreement with the
present lattice calculations \cite{PENE}.
\subsection{Estimate of the bag constant $B_K$}
\nin
We have also estimated the $B_K$-parameter associated to the
$K^0$-$\bar{K}^0
$ mixing,  using the four-quark two-point correlator as
in \cite{BK}. Using the Laplace sum rule (LSR) and adopting
the parametrization of the spectral function in \cite{BK}, we have
obtained the $conservative$ estimate\cite{SNK}:
\beq
B_K \simeq (0.58 \pm 0.22),
\eeq
in good agreement (within the errors) with the FESR result $(0.39\pm 0.10)
$
and with ones from other approaches \cite{PRADES}. However,
our central value is slightly higher than the one from FESR, where the
latter result is
mainly due to the effects of the higher radial excitations in the FESR
analysis which are not under good control. LSR is less sensitive to these
effects due to the exponential factor which suppresses their relative
contributions.
One can also notice that this result from the two-point function sum rule
is more accurate than the one from the three-point function
\cite{3P,CHETY},
which ranges from 0.2 to 1.3, though the
result of \cite{3P} is in good agreement with ours. This inaccuracy can be
intuitively understood from the
relative complexity of the three-point function sum-rule analysis.
\subsection{Estimate of the CP-violation parameters $(\rho,\eta)$}
\nin
We are now ready to discuss the implications of the previous results for
the estimate of the CP-violation parameters $(\rho,\eta)$ defined in
the standard way within the Wolfenstein parametrization \cite{PDG,BURAS}.
Using the previous values of $f_B$, $B_B$ and $B_K$, which are all of
them obtained from a Laplace sum rule analysis, and using the other input
used in \cite{ALI}, one obtains the $best$ fit \cite{SNK}:
\beq
(\rho,\eta) \approx (0.09,0.41),
\eeq
in very good agreement with the expectation in \cite{BURAS} derived from
an alternative method
(see also \cite{PRADES,ALI}). Here, the value of $\rho$ is very sensitive
to
the change of $B_K$ and $f_B$.
\section{The heavy to light exclusive decays}
\subsection{Introduction and notations}
\nin
One can extend the analysis done for the two-point correlator to the
more complicated case of three-point function, in order to study the form
factors related to the $B\rar K^*\gamma$ and $B\rar \rho/\pi$ semileptonic
decays. In so doing, one can consider the generic three-point function:
\beq
V(p,p',q^2)\equiv -\int d^4x~ d^4y ~e^{i(p'x-py)} ~\la 0|{\cal T}
 J_L(x)O(0)J^{\dagger}_B(y)|0\ra ,
\eeq
where $J_L, ~J_B$ are the currents of the light and $B$ mesons; $O$
 is the
weak operator specific for each process (penguin for the $K^* \gamma$,
weak current for the semileptonic); $q \equiv p-p'$ \footnote{
It has to be noticed that we shall use here, like
 in \cite{SN5}-\cite{SN8}, the pseudoscalar current $J_P=(m_u+m_d)\bar{u}(
i
\gamma^5) d$ for describing the pion, where the QCD expression of the
form factor can be deduced from the one in \cite{OVI} by taking $m_c=0$
and
by remarking that the additionnal effect due to the light quark condensate
for $B \rar \pi$ relative to $B \rar D$
 vanishes in the sum rule analysis. In the literature
\cite{COL,BALL2}, the axial-vector
current has been used. However, as it is already well-known in the case of
the
two-point correlator of the axial-vector current, by keeping its
$q_\mu q_\nu$
 part, (which is similarly done in the case of the three-point function)
one obtains the contribution from the $\pi$ plus the $A_1$ mesons but $not
$
the $\pi$ contribution alone. Though, the $A_1$ effect can be
numerically small in the sum rule analysis due
to its higher mass, the mass behaviour of the form factor obtained in this
way differs significantly from the one where the pseudoscalar current has
been used due to the different QCD expressions of the form factor in the
two cases.}.
The vertex obeys the double dispersion
relation :
\beq
V(p^2,p'^2) \simeq
\int_{M_b^2}^{\infty} \frac{ds}{s-p^2-i\epsilon}
\int_{m_L^2}^{\infty} \frac{ds'}{s'-p'^2-i\epsilon}
{}~\frac{1}{\pi^2}~ \mbox{Im}V(s,s')
\eeq
As usual, the QCD part enters in the LHS of the sum rule, while the
experimental observables can be introduced through the spectral function
after the introduction of the intermediate states. The improvement of the
dispersion relation can be done in the way discussed
previously for the two-point
function. In the case of the heavy to light transition,
 the only possible
improvement with a good $M_b$ behaviour
at large $M_b$ (convergence of the QCD series) is the so-called
hybrid sum rule (HSR) corresponding to the uses of
the moments for the heavy-quark
channel and to the Laplace for the light one \cite{SNA,SN5}:
\beq
{\cal H} (n, \tau') =\frac{1}{\pi^2}
\int_{M^2_b}^\infty \frac{ds}{s^{n+1}}
\int_0^\infty ds'~e^{-\tau' s'}~\mbox{Im}V(s,s').
\eeq
The different form factors entering the previous
processes  are defined as:
\bea
 \la\rho(p')\ve \bar u \gamma_\mu (1-\gamma_5) b \ve B(p)\ra
=(M_B+M_\rho)A_1
\epsilon^*_\mu - \nnb \\
 \frac{A_2}{M_B+M_\rho}\epsilon^*p'(p+p')_\mu
+\frac{2V}{M_B+M_\rho} \epsilon_{\mu \nu \rho \sigma}p^\rho p'^\sigma ,
  \nnb \\
 \la\pi(p')\ve
\bar u\gamma_\mu b\ve B(p)\ra = f_+(p+p')_{\mu}+f_-(p-p')_\mu ,
 \nnb \\
\eea
and:
\bea
&&\la \rho(p')\ve \bar s \sigma_{\mu \nu}\ga
\frac{1+\gamma_5}{2}\dr q^\nu b\ve B(p)\ra \nnb \\
&&=i\epsilon_{\mu \nu \rho \sigma}\epsilon^{*\nu}p^\rho p'^\sigma
F^{B\rar\rho}_1 +
\nnb \\
&& \aga \epsilon^*_\mu(M^2_B-M^2_{\rho})-\epsilon^*q(p+p')_{\mu}
\adr \frac{F^{B\rar \rho}_1}{2}. \nnb \\
&&
\eea
\subsection{$q^2$ and $M_b$-behaviours of the form factors}
\nin
We have studied analytically the previous form factors
\cite{SN6}-{\cite{SN8}.
We found that they are dominated, for $M_b \rar \infty$, by the
effect of the light-quark condensate, which dictates the $M_b$ behaviour
of the form factors to be typically of the form:
\beq
F(0) \sim \frac{\la \bar dd \ra}{f_B}\aga 1+\frac{{\cal I}_F}
{M^2_b}\adr,
\eeq
where ${\cal I}_F$ is the integral from the perturbative triangle
graph, which is constant as $t'^2_cE_c/\la \bar dd \ra$ ($t'_c$ and
$E_c$ are the continuum thresholds of the light and $b$ quarks)
for large values of $M_b$. It
indicates that at $q^2=0$ and to leading order in $1/M_b$,
all form factors behave like $\sqrt{M_b}$,
although, in most cases, the coefficient of the $1/M^2_b$ term is large.
The
study of the $q^2$ behaviours of the form factors shows
 that, with the
exception of the $A_1$ form factor, their $q^2$ dependence
is only due to the non-leading (in $1/M_b$) perturbative graph, so that
for $M_b \rar \infty$,
these  form factors remain constant from $q^2=0$ to $q^2_{max}$.
The resulting $M_b$ behaviour at $q^2_{max}$ is the one expected from the
heavy quark symmetry. The numerical
effect of this $q^2$-dependence at finite values of $M_b$ is a polynomial
in $q^2$ (which can be resummed),
which mimics  quite well the usual pole parametrization for a pole mass
of about 6--7 GeV. The situation for the $A_1$ is drastically
different from
the other ones, as here the Wilson coefficient of the $\la \bar dd \ra$
condensate contains a $q^2$ dependence with a $wrong$ sign and reads
\cite{SNA}:
\beq
A_1(q^2) \sim \frac{\la \bar dd \ra}{f_B}\aga 1-\frac{q^2}{M^2_b}
\adr ,
\eeq
which, for $q^2_{max} \equiv (M_B-M_\rho)^2$, gives the expected
behaviour:
\beq
A_1(q^2_{max}) \sim \frac{1}{\sqrt{M_b}}.
\eeq
It should be noticed that the $q^2$ dependence of $A_1$
is in complete contradiction
with the pole behaviour due to its wrong sign. This result explains
the numerical analysis of \cite{BALL2}. One should notice that a recent
phenomenological analysis of the data on the large
longitudinal polarization observed in $B\rar K^*+\Psi$ and a relatively
small ratio of the rates $B\rar K^*+\Psi$ over $B\rar K+\Psi$
\cite{PSI} can only be
simultaneously explained if the $A_1(q^2)$ form factor
decreases \cite{GOURD} as expected from our previous result,
while larger choices of increasing or/and monotonically form factors
fail to explain the data \cite{KAM}.
It is still urgent and important to test this $anomalous$ feature of
the $A_1$-form factor from some other data.
\nin
It should be finally noticed that owing to the overall $1/f_B$ factor, all
form factors have a large $1/M_b$ correction.

\subsection{Numerical estimate of the form factors and decay rates}
\nin
In the numerical analysis, we obtain at $q^2=0$, the value of the $B\rar
K^*\gamma$ form factor \cite{SN6}:
\bea
F_1^{B\rar \rho } &\simeq& 0.27 \pm 0.03,\nnb \\
\frac{F_1^{B\rar K^*}}{F_1^{B\rar \rho}}&\simeq& 1.14 \pm 0.02,
\eea
which leads to the branching ratio $(4.5\pm 1.1)\times 10^{-5}$, in
perfect
agreement with the CLEO data and with the estimate in \cite{ALI}.
One should also notice that, in this case,
the coefficient of the $1/M^2_b$ correction is very large,
which makes dangerous the extrapolation of the
$c$-quark results to higher values of the
quark mass. This extrapolation is often done in some
lattice calculations.

\nin
For the semileptonic decays, QSSR
give a good determination of the ratios of
the form factors with the values for the B-decays\cite{SN5}:
\bea
&&\frac{A_2(0)}{A_1(0)} \simeq \frac{V(0)}{A_1(0)}
\simeq 1.11 \pm 0.01    \nnb  \\
&&\frac{A_1(0)}{F_1^{B\rar \rho}(0)} \simeq 1.18 \pm 0.06    \nnb \\
&&\frac{A_1(0)}{f_+(0)} \simeq 1.40 \pm 0.06,
\eea
though their absolute values are inaccurate \cite{SN5,BALL2}. This is
due to the cancellation of systematic errors in the ratios.
Combining these results with the ``world average" value of $f_+(0)=
0.25 \pm 0.02$ and the one of $F_1^{B\rar \rho}(0)$, one can deduce the
rates:
\bea
&&\Gamma_\pi \simeq (4.3\pm 0.7)
|V_{ub}|^2 \times 10^{12}~\mbox{s}^{-1} \nnb \\
&&\Gamma_\rho /\Gamma_\pi \simeq 0.9 \pm 0.2
\eea
These results are quite precise and
indicate the possibility to reach $V_{ub}$ with a good accuracy
from the exclusive modes.
One should notice
here, mainly because of the non-pole behaviour of $A^B_1$,
the ratio between the widths into $\rho$ and into
$\pi$ is about 1, while in different pole models,
 it ranges from 3 to 10. Recent data on $B\rar K(K^*)+ \Psi(\Psi')$decays
\cite{PSI} favour this result. For the
asymmetry, one obtains a large negative value of $\alpha$, contrary to
the case of the pole models.
\subsection{$SU(3)$ breaking in $\bar B/D \rar Kl\bar\nu$ and
   determination of $V_{cd}/V_{cs}$}
\nin
We extend the previous analysis for the estimate of the $SU(3)$ breaking
in the ratio of the form factors:
\beq
 R_P\equiv f_{+}^{P\rar K}(0) /f_{+}^{P\rar \pi}(0),
\eeq
where $P\equiv \bar B,~D$.
As mentioned before, we use
the hybrid moments for the $B$ and the double exponential sum rules for
the D. The analytic expression of $R_P$ is given in
 \cite{SN8}, which leads to the numerical result:
\beq
R_B = 1.007 \pm 0.020 ~~~~~~~R_D=1.102\pm 0.007,
\eeq
where one should notice that for $M_b \rar \infty$, the SU(3) breaking
vanishes, while its size at finite mass is typically of the same  sign
and magnitude
as the one of $f_{D_s}$ or of the $B\rar K^*\gamma$ discussed before.
What is more surprising is the fact that using the previous value of $R_D
$ with the present value of CLEO data \cite{CLEO}:
\beq
\frac{Br(D^+\rar \pi^0 l\nu)}
{Br(D^+\rar \bar{K}^0 l\nu)}= (8.5 \pm 2.7 \pm 1.4)\%,
\eeq
one deduces \footnote{The old MARKIII data \cite{MARK}
would imply a value $0.25 \pm 0.15$.}:
\beq
V_{cd}/V_{cs}=0.322\pm 0.056,
\eeq
Using $|V_{cd}|=0.204 \pm 0.017$ from PDG \cite{PDG}, one then obtains:
\beq
V_{cs}=0.63\pm 0.12.
\eeq
We can also determine directly the absolute value of the $D\rar K$ form
factor. We obtain:
\beq
f_{+}^{D\rar K}(0)\simeq 0.80 \pm 0.16,
\eeq
which used into the CLEOII data \cite{PDG}:
\beq
\Big{|}f_{+}^{D\rar K}(0)\Big{|}^2\Big{|}V_{cs}\Big{|}^2 \simeq 0.495
\pm 0.036,
\eeq
leads to:
\beq
V_{cs}=0.88\pm 0.18.
\eeq
The average of our two determinations is:
\beq
V_{cs}=0.71\pm 0.10,
\eeq
which needs a confirmation of the CLEOII data. One can compare this value
with the one quoted by PDG94 \cite{PDG}. We expect that the most
reliable result is the lower bound derived from Eq. (70) and from $
f_{+}^{D\rar K}(0)\leq 1$, which is:
\beq
V_{cs}\geq 0.62,
\eeq
while the value $V_{cs}\simeq 1.01 \pm 0.18$ quoted there is related to
the
choice $f_{+}^{D\rar K}(0)\simeq 0.70 \pm 0.1$.
\section{Slope of the Isgur--Wise function and determination of
$V_{cb}$}
\nin
Let me now
discuss  the slope of the Isgur--Wise function. Taron--de Rafael
\cite{TARON} have exploited
the analyticity of the elastic $b$-number form factor
$F$ defined as:
\beq
\la B(p')|\bar b \gamma^\mu b |B(b)\ra =(p+p')^\mu F(q^2),
\eeq
which is normalized as $F(0)=1$
in the large mass limit $M_B \simeq M_D$. Using the positivity of the
vector spectral function and a mapping in order to get a bound on the
slope of $F$
outside the physical cut, they obtained a rigorous but weak bound:
\beq
F'(vv'=1) \geq -6.
\eeq
Including the effects of the $\Upsilon$ states below $\bar BB$ thresholds
by assuming that the $\Upsilon \bar BB$ couplings are
of the order of 1, the
bound becomes stronger:
\beq
F'(vv'=1) \geq -1.5.
\eeq
Using QSSR, we can estimate the part
of these couplings entering in the elastic form factor.
We obtain the value of their sum \cite{SN7}:
\beq
\sum g_{\Upsilon \bar BB} \simeq 0.34 \pm 0.02.
\eeq
In order to be conservative, we have multiplied the previous estimate
by a factor 3 larger. We thus obtain the improved bound:
\beq
F'(vv'=1) \geq -1.34,
\eeq
but the gain over the previous one is not much. Using the
relation of the form factor with the slope of the Isgur--Wise function,
which differs by $-16/75 \log \alpha_s (M_b)$ \cite{FALK},
one can deduce the final bound:
\beq
\zeta'(1) \geq -1.04.
\eeq
However, one can also use the QSSR expression of the Isgur--Wise function
from vertex sum rules \cite{NEU} in order to extract
the slope $analytically$. To leading order in 1/M,
the $physical$ IW function reads:
\bea
\zeta_{phys}(y\equiv vv')&=& \ga \frac{2}{1+y} \dr ^2
\Bigg \{
1 +\frac{\alpha_s}{\pi}
f(y) \nnb \\
&-&\la \bar dd \ra \tau^3 g(y) +
\la \alpha_sG^2 \ra \tau^4 h(y) \nnb \\
&+&g\la \bar dGd \ra \tau^5 k(y) \Bigg \},
\nnb \\
&&
\eea
where $\tau$ is the Laplace sum rule variable
and $f,~ h$ and $k$ are analytic functions of $y$. From this expression,
one
can derive the analytic form of the slope \cite{SN7}:
\beq
\zeta'_{phys}(y=1) \simeq -1 + \delta_{pert} + \delta_{NP},
\eeq
where at the $\tau$-stability region:
$
\delta_{pert} \simeq -\delta_{NP} \simeq -0.04,
$
which shows the near-cancellation of the non-leading corrections.
Adding a generous $50 \%$
error of 0.02 for the correction terms, we finally deduce
the {\it leading order result in 1/M}:
\beq
  \zeta'_{phys}(y=1) \simeq -1 \pm 0.02.
\eeq
Using this result in different existing model parametrizations,
we deduce the value of the mixing angle, {\it to leading order in 1/M}:
\beq
V_{cb} \simeq \ga\frac{1.48~\mbox{ps}}{\tau_b}\dr^{1/2}\times
(37.3 \pm 1.2 \pm 1.4)\times 10^{-3},
\eeq
where the first error comes from the data and the second one from the
model-dependence.

\nin
 Let us now discuss the effects due to the
$1/M$ corrections. In so doing, we combine the predicted value of the
form factor\footnote{We have taken a compromise value between the ones
in \cite{VAIN}.} $0.91\pm 0.03$ at y=1, with the one $0.53 \pm 0.09$
\footnote{This value is just on top of the CLEO data \cite{ROSS}.}
from the sum rule in the full theory
(without a 1/M-expansion) at $q^2=0$ \cite{SN5}. The model
dependence of the analysis enters through the concavity of
the form factor between these two extremal boundaries. We use a linear
parametrization:
\beq
\zeta = \zeta_0 +\zeta'(y-1),
\eeq
which is also supported by the CLEO data \cite{ROSS}. Then,
we can deduce the slope:
\beq
 \zeta' \simeq -( 0.76 \pm 0.2).
\eeq
 It indicates that the $1/M$
correction tends also to decrease the value of $\zeta'$,
which implies that, for
larger values of $y$ where the data are more accurate,
the increase of $V_{cb}$ is weaker (+ 3.7$\%$) than the
one at $y=1$. This leads to the $final$ estimate:
\beq
V_{cb} \simeq \ga\frac{1.48~\mbox{ps}}{\tau_b}\dr^{1/2}
\times (38.8 \pm 1.2 \pm 1.5 \pm 1.5)\times 10^{-3},
\eeq
where the new last error is induced by the error from the slope, while the
model dependence only brings a relatively small error. Our results
for the slope and for $V_{cb}$ are in good
agreement with the new CLEO data \cite{ROSS}.
\nin
However, despite its model dependence, we expect that
the result for $V_{cb}$ is more precise than the one obtained by
exploiting
the value of the Isgur-Wise function at $y=1$ \cite{NEU2}, where the data
near this point are quite inaccurate.
\nin
It also shows that the value from the exclusive channels
is slightly lower than the present result from the inclusive mode
\cite{VAIN},
which is largely affected by the large uncertainty
in the quark-mass definition and in the heavy quark
kinetic energy entering into the inclusive process.
\section{Properties of the hybrids and $B_c$-like hadrons}
\nin
Let me conclude this talk by shortly discussing the masses of the
hybrid $\bar QGQ$ and the mass and decays of the $B_c$like-hadrons.
\subsection{The hybrids}
\nin
Hybrid mesons are interesting because of
 their exotic quantum numbers. Moreover,
it is not clear if these states are true resonances or if they only
manifest themselves
as a wide continuum instead.
\nin
The lowest $\bar cGc$ states appear to be a
$1^{+-}$ of mass around 4.1 GeV \cite{SNB}.
\nin
The available sum-rule analysis of
the $1^{-+}$ state is not very conclusive due to the absence of stability
for this channel. However, the analysis indicates
that the spin-one states are in the range 4.1--4.7 GeV, in agreement with
the predictions from alternative methods \cite{CLOSE}.
Their characteristic
decays should occur via the $\eta'~ U(1)$-like particle
produced together with a
$\psi$ or an $\eta_c$. However, the phase-space
suppression can be quite important for these reactions.
\nin
The sum rule predicts
that the $0^{--},~0^{++}$ $\bar c Gc$ states are in
the range 5--5.7 GeV, i.e.
about 1 GeV above the spin one.
\nin
Intensive searches of these particles in the
next $\tau$-charm and $B$ factories are an alternative test of our idea
about the confinement of QCD.
\subsection{The $B_c$-like hadrons}

\nin
We have estimated the $B_c$-meson mass and coupling by combining the
results
from potential models and QSSR \cite{BC}. We predict the spectra of the
$B_c$-like hadrons from potential models:
\bea
M_{B_c}&=& (6255 \pm 20) ~\mbox{MeV},\nnb \\
M_{B^*_c}&=&(6330 \pm 20)~\mbox{MeV},\nnb  \\
M_{\Lambda(bcu)}&=& (6.93\pm 0.05)~\mbox{GeV},\nnb \\
M_{\Omega(bcs)}&=&(7.00 \pm 0.05)~\mbox{GeV}, \nnb \\
M_{\Xi^*(ccu)}&=&(3.63 \pm 0.05)~\mbox{GeV},\nnb \\
M_{\Xi^*(bbu)}&=&(10.21 \pm 0.05)~\mbox{GeV},
\eea
which are consistent with, but more precise than,
the sum-rule results.
\nin
The decay constant of the $B_c$
meson is better determined from QSSR. The average of the sum rules with
the
potential model results reads:
\beq
f_{B_c} \simeq (2.94 \pm 0.12)f_\pi ,
\eeq
which leads to the leptonic decay rate into $\tau \nu_\tau $ of
about $(3.0 \pm 0.4)\times (V_{cb}/0.037)^2 \times 10^{10}~ \mbox{s}^{-1}$

\nin
We have also studied the semileptonic decay of the $B_c$ mesons
and the $q^2$-dependence of the form factors.
We found that, in all cases, the QCD predictions
increase faster than the usual pole dominance ones.
The $q^2$-behaviour of the form factor can be fitted with an
effective pole mass of about 4.1--4.6 GeV
instead of the 6.3 GeV expected from a pole model.
Basically, we also found that
each exclusive channel has almost
the same rate which is about 1/3 of the leptonic
one, a result which is in contradiction with the potential model one
\cite{LU}.
\nin
Detection  of these particles in the next
$B$-factory machine will then serve
as a stringent
test of the results from the potential models and QSSR
analysis. The previous analysis is at present extended to the case of
the $B^*_c$ meson \cite{PETER}.
\section{Conclusion}
\nin
We have shortly presented different results from QCD spectral sum rules
in the heavy-quark sector,
which are useful for further theoretical studies
and complement the results from
lattice calculations or/and heavy-quark symmetry.
{}From the experimental point of view,
QSSR predictions agree with available data,
but they also lead to some new features,
 which need to be tested in forthcoming
experiments.
\vfill \eject
\section*{Acknowledgement}
\nin
It is a pleasure to thank all the organizers for
their efficient efforts in making a successful and enjoyable meeting.

\end{document}